\documentclass[12pt,fleqn]{iopart}

\expandafter\let\csname equation*\endcsname=\relax
\expandafter\let\csname endequation*\endcsname=\relax
\usepackage{amsmath}

\usepackage{iopams}

\usepackage{bbm}
\usepackage{color}
\usepackage{graphicx}
\usepackage{epsfig}
\usepackage{yfonts}
\usepackage{xspace}
\newcommand{\abs}[1]{\vert #1 \vert}
\newcommand{\D}{\rmd}
\newcommand{\bleq}{\mathrel{\phantom{=}}}
\newcommand{\norm}[1]{\Vert #1 \Vert}
\newcommand{\SNE}{Schr\"odinger-Newton equation\xspace}
\newcommand{\reals}{\mathbb{R}}
\newcommand{\complex}{\mathbb{C}}
 
\newcommand{\Vg}{V_{\rm g}}
\newcommand{\Vo}{V_{\rm other}}
\newcommand{\Ug}{U_{\rm g}}
\newcommand{\Intg}{I_{\rm g}}
\newcommand{\erf}{\mathrm{erf}}
\newcommand{\eps}{\varepsilon}

\renewcommand{\vec}[1]{\mathrm{\mathbf{#1}}}

\newcommand{\rwf}{\widetilde{\rho}}
\newcommand{\rob}{\rho_c}

\newtheorem{Assumption}{Assumption}

\begin{document}
\ifx\href\undefined\else\hypersetup{linktocpage=true}\fi

\title{Centre-of-mass motion in multi-particle
      Schr\"odinger-Newton dynamics}

\author{Domenico Giulini$^{1,2}$ and Andr\'e Gro{\ss}ardt$^{1,3}$}

\address{$^1$ Center of Applied Space Technology and Microgravity\\
        University of Bremen, Am Fallturm 1, D-28359 Bremen, Germany}

\address{$^2$ Institute for Theoretical Physics\\
        Leibniz University Hannover, Appelstr. 2, D-30167 Hannover, Germany}
        
\address{$^3$ Department of Physics\\
        University of Trieste,
        Strada Costiera 11, I-34151 Miramare-Trieste, Italy}

\eads{\mailto{giulini@itp.uni-hannover.de}, \mailto{andre.grossardt@ts.infn.it}}

\begin{abstract}
\noindent
We investigate the implication of the non-linear and non-local 
multi-particle Schr\"odinger-Newton equation for the motion of 
the mass centre of an extended multi-particle object, giving 
self-contained and comprehensible derivations. In particular, 
we discuss two opposite limiting cases. In the first case, the 
width of the centre-of-mass wave packet is assumed much larger 
than the actual extent of the object, in the second case it is 
assumed much smaller. Both cases result in non-linear deviations 
from ordinary free Schr\"odinger evolution for the centre of mass. 
On a general conceptual level we include some discussion in 
order to clarify the physical basis and intention for studying 
the Schr\"odinger-Newton equation.  
\end{abstract}

\pacs{03.65.-w, 04.60.-m}
\ams{35Q40}

\section{Introduction}
How does a quantum system in a non-classical state gravitate? 
There is no unanimously accepted answer to this seemingly obvious 
question. If we assume that gravity is fundamentally quantum, 
as most physicists assume, the fairest answer is simply that we 
don't know. If gravity stays fundamentally classical, a perhaps 
less likely but not altogether outrageous possibility 
\cite{Rosenfeld:1963,Carlip:2008}, we also don't know; but 
we can guess.
One such guess is that semi-classical gravity stays valid, 
beyond the realm it would be meant for if gravity were quantum
\cite{Rosenfeld:1963,Carlip:2008}. Semi-classical gravity in 
that extended sense is the theory which we wish to pursue 
in this paper. Since eventually we are aiming for the 
characterisation of experimentally testable consequences 
of such gravitational self-interaction through matter-wave 
interferometry, we focus attention on the centre-of-mass 
motion. 

Note that by ``quantum system'' we refer to the possibility for 
the system to assume states which have no classical counterpart,
like superpositions of spatially localised states. We are not primarily
interested in matter under extreme conditions (energy,
pressure, etc.). Rather we are interested in ordinary laboratory 
matter described by non-relativistic Quantum Mechanics, whose 
states will source a classical gravitational field according to 
semi-classical equations. Eventually we are interested in the 
question concerning the range of validity of such equations. 
Since we do not exclude the possibility that gravity might stay 
classical at the most fundamental level, we explicitly leave 
open the possibility that these equations stay valid even for 
strongly fluctuating states of matter. 

Now, if we assume that a one-particle state $\psi$ gravitates 
like a classical mass density $\rwf(\vec x)=m\abs{\psi(\vec x)}^2$, 
we immediately get the coupled equations (neglecting other external 
potentials for simplicity) 
\begin{subequations}
\label{eq:SNE-Sys}
\begin{alignat}{1}
\label{eq:SNE-Sys-a}
\mathrm{i}\hbar\partial_t\psi(t;\vec x)
&\,=\,\left(-\frac{\hbar^2}{2m}\Delta+\Vg(t;\vec x)\right)\,\psi(t;\vec x)\,,\\
\label{eq:SNE-Sys-b}
\Delta\Vg(t;\vec x)
&\,=\,4\pi\,G\,m^2\,\abs{\psi(t;\vec x)}^2\,.
\end{alignat}
\end{subequations}
These equations are known as the (one-particle) \emph{Schr\"odinger-Newton system}. 
This system can be transformed into a single, non-liner and non-local 
equation for $\psi$ by first solving \eqref{eq:SNE-Sys-b} 
with boundary condition that $\phi$ be zero at spatial infinity, which 
leads to 
\begin{equation}
\label{eq:GravPot}
\Vg(t;\vec x)=-Gm^2\int\frac{\abs{\psi(t;\vec x')}^2}{\norm{\vec x-\vec x'}}\,\D^3\vec x'\,.
\end{equation}
Inserting \eqref{eq:GravPot} into \eqref{eq:SNE-Sys-a} results in the 
one-particle \emph{Schr\"odinger-Newton equation}:
\begin{equation}
\label{eq:SNE}
\mathrm{i}\hbar\partial_t\psi(t;\vec x)
\,=\,\left(-\frac{\hbar^2}{2m}\Delta
-Gm^2\int\frac{\abs{\psi(t;\vec x')}^2}{\norm{\vec x-\vec x'}}\,\D^3\vec x'\right)\,\psi(t;\vec x)\,.
\end{equation}
Concerning the theoretical foundation of \eqref{eq:SNE}, the non-linear 
self interaction should essentially be seen as a falsifiable hypothesis 
on the gravitational interaction of matter fields, where the reach of 
this hypothesis delicately depends on the kind of ``fields'' it is 
supposed to cover. For example, \eqref{eq:SNE} has been shown to 
follow in a suitable non-relativistic limit from the  Einstein--Klein-Gordon or Einstein-Dirac systems~\cite{Giulini.Grossardt:2012}, i.\,e., 
systems where the energy-momentum tensor $T_{\mu\nu}$ on the right-hand 
side of Einstein's equations,
\begin{equation}
\label{eq:EinsteinEq}
R_{\mu\nu}-\tfrac{1}{2}g_{\mu\nu}R=\frac{8\pi G}{c^4}\,T_{\mu\nu}\,,
\end{equation}
is built from classical Klein-Gordon or classical Dirac fields.
Such an expression for $T_{\mu\nu}$ results from the expectation 
value $\langle\psi\vert{\hat T}_{\mu\nu}\vert\psi\rangle$ in 
Quantum-Field Theory, where $\psi$ labels the amplitude 
(wave function) of a one-particle state, ${\hat T}_{\mu\nu}$ is the 
operator-valued energy-momentum tensor which has been suitably
regularised.\footnote{Defining a suitably regularised 
energy-momentum operator of a quantum field in curved space-time 
is a non-trivial issue; see, e.\,g.,~\cite{Wald:1994}.} 
The non-relativistic limit is then simply the (regularised) 
mass density operator whose expectation value in a one-particle 
state is $m \abs{\psi}^2$; see e.\,g.~\cite{Anastopoulos.Hu:2014-b}. 

Now, if we believe that there exists an underlying quantum 
theory of gravity of which the semi-classical Einstein 
equation \eqref{eq:EinsteinEq} with $T_{\mu\nu}$ replaced by 
$\langle\psi\vert{\hat T}_{\mu\nu}\vert\psi\rangle$ is only 
an approximation, then this will clearly only make sense in 
situations where the source-field for gravity, which is an 
operator, may be replaced by its mean-field approximation. 
This is the case in many-particle situations, i.\,e., where 
$\psi$ is a many-particle amplitude, and then only in the limit as 
the particle number tends to infinity. From that perspective 
it would make little sense to use one-particle expectation values 
on the right hand side of Einstein's equation, for their 
associated classical gravitational field according to 
\eqref{eq:EinsteinEq} will not be any reasonable approximation 
of the (strongly fluctuating) fundamentally quantum 
gravitational field. This has been rightfully stressed 
recently \cite{Anastopoulos.Hu:2014-b,Anastopoulos.Hu:2014-a}.

On the other hand, if we consider the possibility that gravity 
stays fundamentally classical, as we wish to do so here, then 
we are led to contemplate the strict (and not just approximate) 
sourcing of gravitational fields by expectation 
values rather than operators. In this case we \emph{do} get
non-linear self-interactions due to gravity in the equations, 
even for the one-particle amplitudes. Note that it would clearly 
not be proper to regard these amplitudes as classical fields and 
once more (second) quantise them. This is an important 
conceptual point that seems to have caused some confusion 
recently. We will therefore briefly return to this issue at 
the end of section\,\ref{sec:many-particle-sne}. Also recall 
that the often alleged existing evidences, experimental \cite{Page.Geilker:1981} or conceptual \cite{Eppley.Hannah:1977}, 
are generally found inconclusive, e.\,g., \cite{Mattingly:2005,Mattingly:2006,Albers.Kiefer.Reginatto:2008}. 

Taken as a new hypothesis for the gravitational interaction 
of matter, the \SNE\ has attracted much attention in recent years. 
First of all, it raises the challenge to experimentally probe 
the consequences of the non-linear gravitational self-interaction 
term \cite{Salzman.Carlip:2006}. More fundamentally, the verification 
of the existence of this semi-classical self-interaction could 
shed new light on the holy grail of theoretical physics: 
\textgoth{Quantum Gravity} and its alleged necessity; compare \cite{Carlip:2008}.
And even though the original numerical 
estimates made in~\cite{Salzman.Carlip:2006} were too 
optimistic by many orders of magnitude, there is now 
consensus as to the prediction of \eqref{eq:SNE} concerning 
gravity-induced inhibition of quantum-mechanical 
dispersion~\cite{Giulini.Grossardt:2011}. 

However, concerning the current and planned interference 
experiments, it must be stressed that they are  made 
with extended objects, like large molecules or tiny ``nano-spheres''~\cite{HornbergerEtAl:2012}, and that 
the so-called ``large superpositions'' concern only 
the centre-of-mass part of the overall multi-particle 
wave-function. But even if we assume the elementary 
constituents in isolation to obey \eqref{eq:SNE}, there is 
still no obvious reason why the centre of mass of a 
compound object would obey a similar equation. These equations are non-linear and ``separating off'' degrees 
of freedom is not as obvious a procedure as in the linear 
case. The study of this issue is the central concern of this 
paper. For this we start afresh  from a multi-particle 
version of the \SNE.

\section{The many-particle \SNE}
\label{sec:many-particle-sne}
In this paper we consider the $(N+1)$-particle Schr\"odinger-Newton 
equation for a function $\Psi:\reals^{1+3(N+1)}\rightarrow\complex$,
where $3(N+1)$ arguments correspond to the 3 coordinates each of 
$(N+1)$ particles of masses $m_0,m_1,\cdots,m_N$,  and one argument 
is given by the (Newtonian absolute) time $t$. In presence of
non-gravitational 2-body interactions represented by potentials $V_{ab}\bigl(\norm{\vec x_a-\vec x_b}\bigr)$,
where $V_{ab}=V_{ba}$, for the pair labelled by $(ab)$, the 
$(N+1)$-particle Schr\"odinger-Newton equation reads in full glory
\begin{equation}
\label{eq:n-particleSN}
\begin{split}
&\mathrm{i}\hbar\partial_t\Psi(t;\vec x_0,\cdots,\vec x_N)
=\biggl(-\sum_{a=0}^N\frac{\hbar^2}{2m_a}\Delta_a
+\sum_{a=0}^N\sum_{b>a}^N V_{ab}\bigl(\norm{\vec x_a-\vec x_b}\bigr)\biggr.\\
\biggl.&-G\sum_{a=0}^N\sum_{b=0}^N m_am_b\left\{\int\prod_{c=0}^N  \D^3\vec x'_c\right\}
 \frac{\abs{\Psi(t;\vec x'_0,\cdots,\vec x'_N)}^2}{\norm{\vec x_a - \vec x'_b}}
\biggr)\Psi(t;\vec x_0,\cdots \vec x_N)\,.
\end{split}
\end{equation}
Here and in the sequel, we write 
\begin{equation}
\label{eq:DefMeasures}
\D^3\vec x_c:=\D x_c^1\wedge \D x_c^2\wedge \D x_c^3\quad\text{and}\quad
\prod_{c=0}^N \D^3\vec x_c:=\D^3\vec x_0\wedge\cdots\wedge \D^3\vec x_N\,.
\end{equation}

The second, non-linear and non-local potential term is meant to 
represent the gravitational interaction according to a suggestion 
first made in \cite{Diosi:1984}. The structure of this term seems 
rather complicated, but the intuition behind it is fairly simple: 
\begin{Assumption}
Each particle represents a mass distribution  
in physical space that is proportional to its marginal distribution 
derived from $\Psi(t;\vec x_0,\cdots,\vec x_N)$. More precisely, 
the mass distribution represented by the $b$-th particle is  
\begin{equation}
\label{eq:MassDistributionSingleParticle}
\begin{split}
\rwf_b(t;\vec x)
&=m_b \left\{\int\prod_{c=0\atop c\ne b}^N \D^3\vec x_c\right\}\
\abs{\Psi(t;\vec x_0,\cdots,\vec x_{b-1},\vec x,\vec x_{b+1},\cdots,\vec x_N)}^2\\
&=m_b \left\{\int\prod_{c=0}^N \D^3\vec x_c\right\}\
\delta^{(3)}(\vec x-\vec x_b)\,\abs{\Psi(t;\vec x_0,\cdots,\vec x_N)}^2
\end{split}
\end{equation}
\end{Assumption}
\begin{Assumption}
The total gravitational potential $\Phi$ at $\vec x$ in physical 
space is that generated by the sum of the mass distributions 
\eqref{eq:MassDistributionSingleParticle} according to Newtonian 
gravity. More precisely, the Newtonian gravitational potential 
is given by 
\begin{equation}
\label{eq:TotalGravPotential}
\Phi(t;\vec x)=-G\int \D^3\vec x'\frac{\sum_{b=0}^N\rwf_b(t;\vec x')}{\norm{\vec x-\vec x'}}
\end{equation}
\end{Assumption}
\begin{Assumption}
The gravitational contribution $\Vg(\vec x_0,\cdots,\vec x_N)$ 
that enters the Hamiltonian in the multi-particle Schr\"odinger equation 
\begin{equation}
\label{eq:MultiParticleS-Eq}
\begin{split}
\mathrm{i}\hbar\partial_t\psi(t;\vec x_0,\cdots,\vec x_N)
=\biggl(&-\sum_{a=0}^N\frac{\hbar^2}{2m_a}\Delta_a
+\Vo(t;\vec x_0,\cdots,\vec x_N)\biggr.\\
&+\biggl.\Vg(t;\vec x_0,\cdots,\vec x_N)\biggr)\,
\Psi(t;\vec x_0,\cdots,\vec x_N)\,.
\end{split}
\end{equation}
is the sum of the gravitational potential energies of $(N+1)$ 
point-particles (sic!) of masses $m_a$ situated at positions 
$\vec x_a$. More precisely, the total gravitational contribution 
to the Hamiltonian is 
\begin{equation}
\label{eq:GracContribHamiltonian}
\Vg(t;\vec x_0,\cdots,\vec x_N)=\sum_{a=0}^Nm_a\Phi(t;\vec x_a)
\end{equation}
where $\Phi$ is given by \eqref{eq:TotalGravPotential}.
\end{Assumption}
 
Taken together, all three assumptions result in a gravitational 
contribution to the Hamiltonian of 
\begin{equation}
\label{eq:GravHamiltonian}
\Vg(t;\vec x_0,\cdots,\vec x_N)
=-G\sum_{a=0}^N\sum_{b=0}^N m_am_b\left\{\int\prod_{c=0}^N \D^3\vec x'_c\right\}\frac{\abs{\Psi(t;\vec x'_0,\cdots,\vec x'_N)}^2}{\norm{\vec x_a-\vec x'_b}}
\end{equation}
just as in \eqref{eq:n-particleSN}. We note that  \eqref{eq:n-particleSN} can 
be derived from a Lagrangian 
\begin{equation}
\label{eq:TminusU}
L=T-U
\end{equation}
where the kinetic part\footnote{In classical field theory it would be 
physically more natural to regard the second part of the kinetic term 
$\propto\abs{\vec\nabla\Psi}^2$ as part of the potential energy. In Quantum 
Mechanics, however, it represents the kinetic energy of the particles.}, $T$, is  
\begin{equation}
\label{eq:Lagrangian-T}
\begin{split}
T
&=\frac{\mathrm{i}\hbar}{2}\left\{\int\prod_{a=0}^N \D^3\vec x_a\right\}
\Bigl(\bar\Psi\partial_t\Psi-\Psi\partial_t\bar\Psi\Bigr)\\
&\ +\hbar^2\left\{\int\prod_{a=0}^N \D^3\vec x_a\right\}\sum_{b=0}^N\frac{1}{m_b}\vec\nabla_b\bar\Psi\cdot\vec\nabla_b\Psi\,.
\end{split}
\end{equation}
Here all functions are taken at the same argument $(t;\vec x_0,\cdots,\vec x_N)$, 
which we suppressed. The potential part, $U$, consists of a sum of two terms. 
The first term represents possibly existent 2-body interactions, like, e.\,g., 
electrostatic energy:
\begin{equation}
\label{eq:Lagrangian-U-2body}
U^{\text{local 2-body}}=\left\{\int\prod_{c=0}^N \D^3\vec x_c\right\}\sum_{a=0}^N\sum_{b>a}^NV_{ab}(t;\vec x_0,\cdots,\vec x_N)
\,\abs{\Psi(t;\vec x_0,\cdots,\vec x_N)}^2
\end{equation}
The second contribution is that of gravity: 
\begin{equation}
\label{eq:Lagrangian-U-gravity}
\begin{split}
U^{\rm grav}=
-\frac{G}{2}&\left\{\int\prod_{c=0}^N \D^3\vec x_c\right\}
\left\{\int\prod_{d=0}^N \D^3\vec x'_d\right\}\sum_{a=0}^N\sum_{b=0}^N m_am_b\\
&\times
\frac{\abs{\Psi(t;\vec x_0,\cdots,\vec x_N)}^2
\abs{\Psi(t;\vec x'_0,\cdots,\vec x'_N)}^2}{\norm{\vec x_a-\vec x'_b}}\\
=-\frac{G}{2}&\sum_{a=0}^N\sum_{b=0}^N \int \D^3\vec x\int \D^3\vec x'\quad
\frac{\rwf_a(\vec x)\,\rwf_b(\vec x')}{\norm{\vec x-\vec x'}}
\end{split}
\end{equation}
The last line shows that the gravitational energy is just the usual 
binding energy of $(N+1)$ lumps of matter distributed in physical 
space according to \eqref{eq:MassDistributionSingleParticle}. 
Note that the sum not only contains the energies for the mutual 
interactions between the lumps, but also the self-energy of each 
lump. The latter are represented by the diagonal terms in the double 
sum, i.\,e. the terms where $a=b$. These self-energy contributions would 
diverge for pointlike mass distributions, i.\,e. if 
$\rwf_a(\vec x)=m_a\delta^{(3)}(\vec x-\vec x_a)$, as in the 
case of electrostatic interaction (see below). Here, however, the 
hypotheses underlying the three assumptions above imply that 
gravitationally the particles interact differently, resulting in 
finite self-energies. Because of these self-energies we already 
obtain a modification of the ordinary Schr\"odinger equation in 
the one-particle case, which is just given by~\eqref{eq:SNE}. 
Explicit expressions for the double integrals over 
$\rwf_a(\vec x)\rwf_b(\vec x')/\norm{\vec x-\vec x'}$ can, e.\,g., 
be found in \cite{Iwe:1982} for some special cases where $\rwf_a$ 
and $\rwf_b$ are spherically symmetric. 

Finally we wish to come back to the fundamental issue already 
touched upon in the introduction, namely of how to relate 
the interaction term \eqref{eq:Lagrangian-U-gravity} to known 
physics as currently understood. As already emphasised in the 
context of \eqref{eq:SNE}, i.\,e. for just one particle, the 
gravitational interaction contains self-energy contributions. 
In the multi-particle scheme they just correspond to the diagonal 
terms  $a=b$ in \eqref{eq:Lagrangian-U-gravity}. These terms 
are certainly finite for locally bounded $\rwf_a$. 

This would clearly not be the case in a standard quantum 
field-theoretic treatment, like QED, outside the mean-field 
limit. In non-relativistic Quantum Field Theory the interaction 
Hamiltonian would be a double integral over
$\Psi^\dagger(\vec x)\Psi(\vec x)\Psi^\dagger(\vec x')\Psi(\vec x')/\norm{\vec x-\vec x'}$, where $\Psi$ is the (non-relativistic) field operator.
(See, e.\,g.,  chpater\,11 of \cite{Grawert:QM} for a text-book account 
of non-relativistic QFT.) This term will lead to divergent self 
energies, which one renormalises through normal ordering, and 
pointwise Coulomb interactions of pairs. This is just the known 
and accepted strategy followed in deriving the multi-particle 
Schr\"odinger equation for charged point-particles from QED. 
This procedure has a long history. In fact, it can already be found 
in the Appendix of Heisenberg's 1929 Chicago lectures 
\cite{Heisenberg:1949} on Quantum Mechanics.

It has therefore been frequently complained that the \SNE\ 
does \emph{not} follow from ``known physics'' 
\cite{Christian:1997,Adler:2007,Anastopoulos.Hu:2014-b,Anastopoulos.Hu:2014-a}.
This is true, of course. But note that this does not imply the 
sharper argument according to which the \SNE\ even contradicts 
known physics. Such sharper arguments usually beg the question 
by assuming some form of quantum gravity to exist. But this 
hypothetical theory is not yet part of ''known physics'' either, 
and may never be! Similarly, by rough analogy of the classical 
fields in gravity and electromagnetism, the \SNE\ is sometimes 
argued to contradict known physics because the analogous non-linear 
``Schr\"odinger-Coulomb'' equation yields obvious nonsense, 
like a grossly distorted energy spectrum for hydrogen. In fact, 
this has already been observed in 1927 by Schr\"odinger who 
wondered about this factual contradiction with what he described 
as a natural demand (self coupling) from a classical 
field-theoretic point of view~\cite{Schroedinger:1927}. 
Heisenberg in his 1929 lectures also makes this observation which 
he takes as irrefutable evidence for the need to (second) quantise 
the Schr\"odinger field, thereby turning a non-linear ``classical'' 
field theory into a linear quantum version of it. 

To say it once more: all this is only an argument against 
the \SNE\ provided we assume an underlying theory of quantum 
gravity to exist and whose effective low energy approximation can 
be dealt with in full analogy to, say, QED. But our attitude 
here is different! What we have is a hypothesis that is 
essentially based on the assumption that gravity behaves 
differently as regards its coupling to matter and, in particular, 
its need for quantisation. The interesting aspect of this 
is that it gives rise to potentially observable consequences 
that render this hypothesis falsifiable. 

\section{Centre-of-mass coordinates}
\label{sec:com-coordinates}
Instead of the $(N+1)$ positions $\vec x_a$, $a=0,\cdots,N$,
in absolute space, we introduce the centre of mass and $N$ positions 
relative to it. We write 
\begin{equation}
\label{eq:TotalMass}
M:=\sum_{a=0}^Nm_a
\end{equation} 
for the total mass and adopt the convention that greek indices
$\alpha,\beta,\cdots$ take values in $\{1,\cdots,N\}$, in 
contrast to latin indices $a,b,\cdots$, which we already agreed to 
take values in  $\{0,1,\cdots,N\}$. The centre-of-mass and the relative
coordinates of the $N$ particles labelled by $1,\cdots,N$ are 
given by (thereby distinguishing the particle labelled by $0$)
\begin{subequations}
\label{eq:com-coordinates} 
\begin{alignat}{1}
\label{eq:com-coordinates-a} 
&\vec c
\,:=\,\frac{1}{M}\sum_{a=0}^N m_a\,\vec x_a
=\frac{m_0}{M}\vec x_0
+\sum_{\beta=1}^N \frac{m_\beta}{M}\vec x_\beta\,,\\
\label{eq:com-coordinates-b} 
&\vec r_\alpha\,:=\,\vec x_\alpha-\vec c
=-\frac{m_0}{M}\,\vec x_0+\sum_{\beta=1}^N\left(\delta_{\alpha\beta}
-\frac{m_\beta}{M}\right)\vec x_\beta
\end{alignat}
\end{subequations}
The inverse transformation is obtained by simply solving \eqref{eq:com-coordinates} for $\vec c$ and $\vec r_\alpha$: 
\begin{subequations}
\label{eq:com-coordinates-inverse} 
\begin{alignat}{1}
\label{eq:com-coordinates-inverse-a} 
&\vec x_0=\vec c-\sum_{\beta=1}^N \frac{m_\beta}{m_0}\,\vec r_\beta\,,\\
\label{eq:com-coordinates-inverse-b} 
&\vec x_\alpha=\vec c+\vec r_\alpha\,.
\end{alignat}
\end{subequations} 
All this may be written in a self-explanatory $(1+N)$ split matrix 
form
\begin{alignat}{2}
\label{eq:com-coordinates-matrixform-1}
\begin{pmatrix}
\vec c\\
\vec r_\alpha
\end{pmatrix}
&=
\begin{pmatrix}
\frac{m_0}{M}&\frac{m_\beta}{M}\\
-\frac{m_0}{M}&\delta_{\alpha\beta}-\frac{m_\beta}{M}
\end{pmatrix}
\begin{pmatrix}
\vec x_0\\
\vec x_\beta
\end{pmatrix}\,,
\\
\label{eq:com-coordinates-matrixform-2}
\begin{pmatrix}
\vec x_0\\
\vec x_\alpha
\end{pmatrix}
&=
\begin{pmatrix}
1&-\frac{m_\beta}{m_0}\\
1&\delta_{\alpha\beta}
\end{pmatrix}
\begin{pmatrix}
\vec c\\
\vec r_\beta
\end{pmatrix}\,.
\end{alignat}

For the wedge product of the $(N+1)$ 1-forms
$\D x^1_a$ for $a=0,1,\cdots,N$ we easily get from \eqref{eq:com-coordinates-inverse}
\begin{equation}
\label{eq:com-measure}
\begin{split}
\D x^1_0\wedge\cdots\wedge \D x^1_N
&=\Biggl(\D c^1-\sum_{\beta=1}^N\frac{m_\beta}{m_0}\D r^1_\beta\Biggr)
\wedge 
\bigl(\D c^1+\D r^1_1\bigr)\wedge\cdots\wedge\bigl(\D c^1+\D r^1_N\bigr)\\
&=\frac{M}{m_0}\bigl(\D c^1\wedge \D r^1_1\wedge\cdots\wedge \D r^1_N\bigr)\,.
\end{split}
\end{equation}
Hence, writing $\D^3\vec x_\alpha:=\D x^1_\alpha\wedge \D x^2_\alpha\wedge \D x^3_\alpha$
and $\prod_{\alpha=1}^N$ for the $N$-fold wedge product, we have   
\begin{equation}
\label{eq:WedgeProducts}
 \prod_{a=0}^N \D^3\vec x_a=\left(\frac{M}{m_0}\right)^3
 \left(\D^3\vec c\wedge\prod_{\alpha=1}^N\D^3\vec r_\alpha\right)\,.
\end{equation}
Note that the sign changes that may appear in rearranging the 
wedge products on both sides coincide and hence cancel.
From \eqref{eq:WedgeProducts} we can just read off the determinant 
of the Jacobian matrix for the transformation \eqref{eq:com-coordinates-inverse}:
\begin{equation}
\label{eq:JacobianCoordTrans}
\left\vert\frac{\partial(\vec x_0,\vec x_\alpha)}{\partial(\vec c,\vec r_\beta)}\right\vert
:=\det\left\{\frac{\partial(\vec x_0,\vec x_\alpha)}{\partial(\vec c,\vec
r_\beta)}\right\}
=\left(\frac{M}{m_0}\right)^3 \,.
\end{equation} 

Equation \eqref{eq:com-coordinates-inverse} also allows to 
simply rewrite the kinetic-energy metric 
\begin{equation}
\label{eq:DefKinEnergyMetric}
G=\sum_{a=0}^N\sum_{b=0}^NG_{ab}\,\D \vec x_a\otimes \D \vec x_b:=
\sum_{a=0}^Nm_a\,\D \vec x_a\otimes \D \vec x_a
\end{equation}
in terms of the new coordinates:  It is given by 
\begin{equation}
\label{eq:KineticEnergyMetric}
\begin{split}
G
&=m_0\,\left(\D \vec c-\sum_{\alpha=1}^N\frac{m_\alpha}{m_0}\D \vec r_\alpha\right)\otimes\left(\D \vec c-\sum_{\beta=1}^N\frac{m_\beta}{m_0}\D \vec r_\beta\right)\\
&\qquad +\sum_{\alpha=1}^N m_\alpha \bigl(\D \vec c+\D \vec r_\alpha\bigr)\otimes\bigl(\D \vec c+\D \vec r_\alpha\bigr)\\ 
&=M\,\D \vec c\otimes \D \vec c+\sum_{\alpha=1}^N\sum_{\beta=1}^NH_{\alpha\beta}\,\D \vec r_\alpha\otimes \D \vec r_\beta\,.
\end{split}
\end{equation}
The first thing to note is that there are no off-diagonal terms, i.\,e. 
terms involving tensor products between $\D \vec c$ and 
$\D \vec r_\alpha$. This means that the degrees of freedom labelled 
by our $\vec r_a$ coordinates are perpendicular (with respect to the 
kinetic-energy metric) to the centre-of-mass motion. The restriction 
of the kinetic-energy metric to the relative coordinates has the 
components
\begin{equation}
\label{eq:KineticEnergyMetric-Internal}
H_{\alpha\beta}=
\left(\frac{m_\alpha m_\beta}{m_0}+m_\alpha\delta_{\alpha\beta}\right)\,.
\end{equation}  

The determinant of $\{H_{\alpha\beta}\}$ follows from taking 
the determinant of the transformation formula for the
kinetic-energy metric (taking due account of the 3-fold 
multiplicities hidden in the inner products in $\reals^3$)  
\begin{equation}
\label{eq:DetH-1}
\bigl(\det\{G_{ab}\}\bigr)^3\times
\left\vert\frac{\partial(\vec x_0,\vec x_\alpha)}{\partial(\vec c,\vec r_\beta)}\right\vert^2=M^3\times\bigl(\det\{H_{\alpha\beta}\}
\bigr)^3
\end{equation}
which, using \eqref{eq:JacobianCoordTrans} and 
$\det\{G_{ab}\}=\prod_{a=0}^N (m_a/2)$, results in
\begin{equation}
\label{eq:DetH-2}
\det\{H_{\alpha\beta}\}
=\frac{M}{m_0^2}\,\prod_{a=0}^Nm_a\,.
\end{equation}

Finally we consider the inverse of the kinetic-energy metric:
\begin{equation}
\label{eq:InverseKineticEnergyMetric-1}
G^{-1}=\sum_{a=0}^N\sum_{b=0}^N G^{ab}
\frac{\partial}{\partial\vec x_a}\otimes
\frac{\partial}{\partial\vec x_b}
=\sum_{a=0}^N\frac{1}{m_a}\frac{\partial}{\partial\vec x_a}\otimes
\frac{\partial}{\partial\vec x_a}
\end{equation}
Using \eqref{eq:com-coordinates} we have 
\begin{subequations}
\label{eq:DerivTrans}
\begin{alignat}{2}
\label{eq:derivTrans-a}
&\frac{\partial}{\partial\vec x_0}
&&\,=\,\frac{m_0}{M}\left(\frac{\partial}{\partial\vec c}
-\sum_{\alpha=1}^N\frac{\partial}{\partial\vec r_\alpha}\right)\,,\\
\label{eq:derivTrans-b}
&\frac{\partial}{\partial\vec x_\alpha}
&&\,=\,\frac{\partial}{\partial\vec r_\alpha}
+\frac{m_\alpha}{M}\left(\frac{\partial}{\partial\vec c}-\sum_{\beta=1}^N\frac{\partial}{\partial\vec r_\beta}\right)\,.
\end{alignat}
\end{subequations}
Inserting this into \eqref{eq:InverseKineticEnergyMetric-1} we obtain 
the form 
\begin{equation}
\label{eq:InverseKineticEnergyMetric-2}
G^{-1}=\frac{1}{M}
\frac{\partial}{\partial\vec c}\otimes
\frac{\partial}{\partial\vec c}
+\sum_{\alpha=1}^N\sum_{\beta=1}^N H^{\alpha\beta}
\frac{\partial}{\partial\vec r_\alpha}\otimes
\frac{\partial}{\partial\vec r_\beta} \,,
\end{equation}
where $\{H^{\alpha\beta}\}$ is the inverse matrix to $\{H_{\alpha\beta}\}$,
which turns out to be surprisingly simple: 
\begin{equation}
\label{eq:InverseKineticEnergyMetric-3}
H^{\alpha\beta}=\bigl(m^{-1}_\alpha\,\delta_{\alpha\beta}-M^{-1}\bigr)\,.
\end{equation}
In fact, the relation 
$\sum_{\beta=1}^NH_{\alpha\beta}H^{\beta\gamma}=\delta_\alpha^\gamma$
is easily checked from the given expressions.

Note that the kinetic part in \eqref{eq:n-particleSN} is just 
$(-\hbar^2/2)$ times the Laplacian on $\reals^{3(N+1)}$ with 
respect to the kinetic-energy metric. Since $\det(G)$ and $\det (H)$ 
are constant, this Laplacian is just: 
\begin{equation}
\label{eq:KinEnergyLaplacian}
\begin{split}
\Delta_G
&=
\sum_{a=0}^N\sum_{b=0}^NG^{ab}
 \frac{\partial}{\partial\vec x_a}\cdot
 \frac{\partial}{\partial\vec x_b}
=\sum_{a=0}^N\frac{1}{m_a}
 \frac{\partial}{\partial\vec x_a}\cdot
 \frac{\partial}{\partial\vec x_a}\\
&=
\frac{1}{M}\frac{\partial}{\partial\vec c}\cdot
\frac{\partial}{\partial\vec c}
+\sum_{\alpha=1}^N\sum_{\beta=1}^NH^{\alpha\beta}
\frac{\partial}{\partial\vec r_\alpha}\cdot
 \frac{\partial}{\partial\vec r_\beta}\\
&=:\Delta_c+\Delta_r\,.
\end{split}
\end{equation}
Here $\Delta_c$ is the part just involving the three centre-of-mass 
coordinates $\vec c$ and $\Delta_r$ the part involving the derivatives 
with respect to the $3N$ relative coordinates $\vec r_\alpha$. Note that  
there are no terms that mix the derivatives with repect to $\vec c$
and $\vec r_\alpha$, but that $\Delta_r$ mixes any two derivatives 
with respect to $\vec r_\alpha$  due to the second term  on the 
right-hand side of \eqref{eq:InverseKineticEnergyMetric-3}. Clearly, 
a further linear redefinition of the relative coordinates 
$\vec r_\alpha$ could be employed to diagonalise $H_{\alpha\beta}$ 
and $H^{\alpha\beta}$, but that we will not need here.

\section{Schr\"odinger-Newton effect on the centre of mass}
Having introduced the centre-of-mass coordinates, one can consider the possibility
that the wave-function separates into a centre-of-mass and a relative
part,\footnote{Here we include the square-root of the inverse of the Jacobian
determinant \eqref{eq:JacobianCoordTrans} to allow for simultaneous normalisation
to $\norm{\Psi} = \norm{\psi} = \norm{\chi} = 1$, which we imply in the following.}
\begin{equation}
\Psi(t;\vec x_0,\cdots,\vec x_N) = \left(\frac{m_0}{M}\right)^{3/2}
\psi(t;\vec c) \, \chi(t;\vec r_1,\cdots,\vec r_N) \,.
\end{equation}
In order to obtain an independent equation for just the centre-of-mass dynamics
one is, however, left with the necessity to show that equation \eqref{eq:n-particleSN}
also separates for this ansatz.
This is true for the kinetic term, as shown in \eqref{eq:KinEnergyLaplacian},
and it is also obvious for the non-gravitational contribution $V_{ab}$ which
depends on the relative distances, and therefore the relative coordinates, only.

As long as non-gravitational interactions are present these are presumably
much stronger than any gravitational effects. Hence, the latter can be ignored
for the \emph{relative} motion, which leads to a usually complicated but well-known
equation: the ordinary, linear Schr\"odinger equation whose solution becomes
manifest in the inner structure of the present lump of matter.

However, while separating the linear multi-particle Schr\"odinger equation in the
absence of external forces (i.\,e. equation \eqref{eq:n-particleSN} with the gravitational
constant $G$ set to zero) yields a free Schr\"odinger equation for the evolution
of the centre of mass, the $(N+1)$-particle \SNE \eqref{eq:n-particleSN} will
comprise contributions of the gravitational potential to the centre-of-mass motion.
The reason for these to appear is the non-locality of the integral term in the
equation (and not the mere existence of the diagonal term $a=b$
as one could naively assume).

Let us take a closer look at the gravitational potential \eqref{eq:GravHamiltonian}. Using the results
from the previous section, in centre-of-mass coordinates it reads:
\begin{equation}
\label{eq:Vgrav-CM-full}
\begin{split}
\Vg(t;\vec c,\vec r_1,\cdots,\vec r_N)
&= -G \, \int \D^3 \vec c' \abs{\psi(t;\vec c')}^2
\left\{\int\prod_{\gamma=1}^N  \D^3\vec r'_\gamma\right\} \\
&\bleq \times\Bigg[ m_0^2 \frac{\abs{\chi(t;\vec r'_1,\cdots,\vec r'_N)}^2}{\norm{\vec c - \vec c'
- \sum_{\delta=1}^N \frac{m_\delta}{m_0} \left(\vec r_\delta - \vec r'_\delta \right)}} \\
&\bleq + m_0 \sum_{\alpha=1}^N m_\alpha \frac{\abs{\chi(t;\vec r'_1,\cdots,\vec r'_N)}^2}%
{\norm{\vec c - \vec c' - \sum_{\delta=1}^N \frac{m_\delta}{m_0} \vec r_\delta - \vec r'_\alpha}} \\
&\bleq + m_0 \sum_{\alpha=1}^N m_\alpha \frac{\abs{\chi(t;\vec r'_1,\cdots,\vec r'_N)}^2}%
{\norm{\vec c - \vec c' + \vec r_\alpha + \sum_{\delta=1}^N \frac{m_\delta}{m_0} \vec r'_\delta}} \\
&\bleq + \sum_{\alpha=1}^N \sum_{\beta=1}^N m_\alpha m_\beta
\frac{\abs{\chi(t;\vec r'_1,\cdots,\vec r'_N)}^2}{\norm{\vec c - \vec c'
+ \vec r_\alpha - \vec r'_\beta}} \Bigg] \,.
\end{split}
\end{equation}
The $m_0$ dependent terms in the second, third, and fourth line are more intricate than those
in the last line; but they are only $(2N+1)$ out of $(N+1)^2$ terms and therefore can be
neglected for large $N$.\footnote{To be more distinct, assign the label ``0'' to that particle
for which the absolute value of the sum of all $(2N+1)$ terms involving $m_0$ is the smallest.
Then these terms can be estimated against all the others and the error made by their negligence
is of the order $1/N$.}
In this ``large $N$''-approximation only the last double-sum 
in \eqref{eq:Vgrav-CM-full} survives. All  $\vec r'_\gamma$ 
integrations except that where $\gamma=\beta$ can be carried out 
(obtaining the $\beta$-th marginal distributions for 
$\abs{\chi(t;\vec r'_1,\cdots,\vec r'_N)}^2$). Because of the 
remaining integration over $\vec r'_\beta$ we may rename the 
integration variable $\vec r'_\beta\rightarrow\vec r'$, thereby 
removing its fictitious dependence on $\beta$. All this leads to 
the expression  
\begin{equation}
\label{eq:Vgrav-CM}
\begin{split}
\Vg(t;\vec c,\vec r_1,\cdots,\vec r_N)
&= -G \sum_{\alpha=1}^N m_\alpha \int \D^3 \vec c' \int \D^3\vec r'
\frac{\abs{\psi(t;\vec c')}^2\rob(\vec r')}{\norm{\vec c - \vec c'
+ \vec r_\alpha - \vec r'}}  \,,
\end{split}
\end{equation}
where we defined 
\begin{equation}
\label{eq:MatterDensity}
\rob(t;\vec r)
:=\sum_{\beta=1}^N m_\beta \left\{\int\prod_{\gamma=1\atop \gamma\ne \beta}^N \D^3\vec r_\gamma\right\}\
\abs{\chi(t;\vec r_1,\cdots,\vec r_{\beta-1},\vec r,\vec r_{\beta+1},\cdots,\vec r_N)}^2\,.
\end{equation}
This ``relative'' mass distribution is built analogously to 
\eqref{eq:MassDistributionSingleParticle} from the marginal 
distributions, here involving only the relative coordinates
of all but the zeroth particle. In the large $N$ approximation 
this omission of $m_0$ should be neglected and 
$\rob(t;\vec r)$ should be identified as the mass distribution 
relative to the centre of mass. Given a (stationary) solution 
$\chi$ of the Schr\"odinger equation for the relative motion, 
$\rob$ is then simply the mass density of the present lump of 
matter (e.\,g. a molecule) relative to the centre of mass.
Although for the following discussion the time-dependence of 
$\rob$ makes no difference, we will omit it. This may be 
justified by an adiabatic approximation, since the typical 
frequencies involved in the relative motions are much higher 
than the frequencies involved in the centre-of-mass motion. 

Note that the only approximation that entered the derivation of
\eqref{eq:Vgrav-CM} so far is that of large $N$. For the typical 
situations we want to consider, where $N$ is large indeed, this 
will be harmless. However, the analytic form taken by the 
gravitational potential in \eqref{eq:Vgrav-CM} is not yet 
sufficiently simple to allow for a separation into centre-of-mass 
and relative motion. In order to perform such a separation we 
have to get rid of the $\vec r_\alpha$-dependence. This can be 
achieved if further approximations are made, as we shall 
explain now.

\section{Approximation schemes}
\subsection{Wide  wave-functions}
As long as the centre-of-mass wave-function is much wider 
than the extent of the considered object one can assume
that it does not change much over the distance $\vec r_\alpha$, i.\,e.
$\abs{\psi(t;\vec c' + \vec r_\alpha)} \approx \abs{\psi(t;\vec c')}$.
Substituting $\vec c'$ by $\vec c'+\vec r_\alpha$ in  
\eqref{eq:Vgrav-CM} then yields the following potential, 
depending only on the centre-of-mass coordinate:
\begin{equation}
\label{eq:large-wf}
\Vg^{(A)}(t;\vec c) \approx -G M \int \D^3 \vec c' \int \D^3\vec r'\;
\frac{\abs{\psi(t;\vec c')}^2\rob(\vec r')}{\norm{\vec c - \vec c' - \vec r'}}  \,.
\end{equation}
As a result, the equation for the centre of mass is now indeed 
of type \eqref{eq:SNE-Sys} with $\Vg=\Vg^{(A)}$ being given by 
$M$ times the convolution of $\abs{\psi}^2$ with the Newtonian 
gravitational potential for the mass-density $\rob$. 
Case\,A has been further analysed in~\cite{Giulini.Grossardt:2013}.

\subsection{Born-Oppenheimer-Type approximation}
An alternative way to get rid of the dependence of 
\eqref{eq:Vgrav-CM} on the relative coordinates, i.\,e., the 
$\vec r'_\alpha$-dependence on the 
right-hand side, is to just replace $\Vg$ with its expectation 
value in the state $\chi$ of the relative-motion.%
\footnote{We are grateful to Mohammad Bahrami for this idea.}
This procedure corresponds to the Born-Oppenheimer approximation 
in molecular physics where the electronic degrees of freedom 
are averaged over in order to solve the dynamics of the nuclei. 
The justification for this procedure in molecular physics derives 
from the much smaller timescales for the motion of the fast and 
lighter electrons as compared to the slow and heavier nuclei. 
Hence the latter essentially move only according to the averaged 
potential sourced by the electrons. In case of the \SNE the   
justification is formally similar, even though it is clear that 
there is no real material object attached to the centre of mass. 
What matters is that the relative interactions (based on 
electrodynamic forces) are  much stronger than the 
gravitational ones, so that the characteristic frequencies of 
the former greatly exceed those of the latter; compare, e.\,g., 
the discussion in \cite{Yang.etal:2013}.  

Now, the expectation value is easily calculated: 
\begin{align}
\label{eq:born-oppenheimer}
\Vg^{(B)}(t;\vec c) 
&= \left\{\int\prod_{\beta=1}^N \D^3\vec r''_\beta\right\} \abs{\chi(\vec r''_1,\cdots,\vec r''_N)}^2
\;\Vg(t;\vec c,\vec r''_1,\cdots,\vec r''_N) \nonumber\\
&= -G \sum_{\alpha=1}^N m_\alpha \int \D^3 \vec c' \int \D^3\vec r'
\left\{\int\prod_{\beta=1}^N \D^3\vec r''_\beta\right\} \nonumber\\
&\bleq \hspace{8em} \times
\frac{\abs{\psi(t;\vec c')}^2\rob(\vec r')\abs{\chi(\vec r''_1,\cdots,\vec r''_N)}^2}%
{\norm{\vec c - \vec c' - \vec r' + \vec r''_\alpha}}  \nonumber\\
&= -G \int \D^3 \vec c' \int \D^3\vec r' \int \D^3\vec r''\;
\frac{\abs{\psi(t;\vec c')}^2\rob(\vec r')\rob(\vec r'')}{\norm{\vec c - \vec c' - \vec r' + \vec r''}}  \,.
\end{align}
Note that this expression involves one more $\reals^3$
integrations than \eqref{eq:large-wf}. 

\begin{figure}
 \centering
 \includegraphics[bb=28 37 287 183]{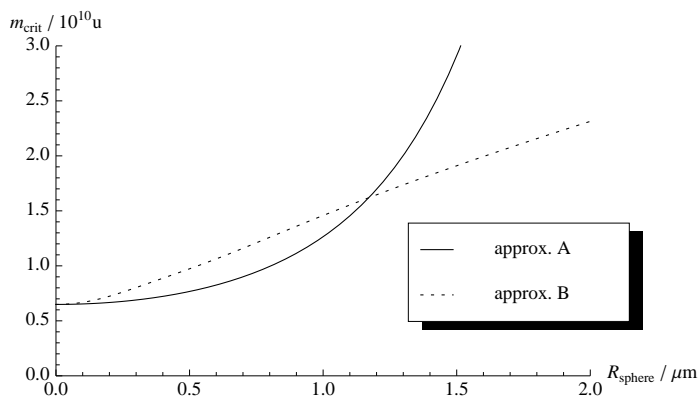}
 \caption{Critical mass for a hollow sphere as indicated by the behaviour of the second moment. We 
 used a wave-packet width of $0.5\,\mu\mathrm{m}$.}
 \label{fig:mcrit-hollow}
\end{figure}

In \cite{Giulini.Grossardt:2013} we studied two simple models 
for the matter density $\rob$: a solid and a hollow sphere. 
The solid-sphere suffers from some peculiar divergence issues 
which we explain in \ref{app:solid-sphere-divergence}
and is also mathematically slightly more difficult to handle 
than the hollow sphere whose radial mass distribution is just a 
$\delta$-function. 
We therefore use the hollow sphere as a model to compare the 
two approximation ans\"atze given above.

While in \cite{Salzman.Carlip:2006,Carlip:2008,Giulini.Grossardt:2011,Giulini.Grossardt:2013}
the expression ``collapse mass'' was used in a rather loosely defined manner,
here we define as critical mass the mass value for which at $t=0$ the second order
time derivative of the second moment $Q(t) = \int \D^3 \vec{c} \, \abs{\vec c}^2 \abs{\psi(t;\vec{c})}^2$
vanishes, i.\,e. $\ddot{Q}(t=0)=0$.
(Note that for a real-valued initial wave-packet the first order time derivative always vanishes.)
For the one-particle \SNE\ and a Gaussian
wave packet of $0.5\,\mu\mathrm{m}$ width this yields a critical mass of $6.5 \times 10^9$\,u
which fits very well with the numerical results obtained in~\cite{Giulini.Grossardt:2011}.

For the hollow sphere we then obtain the analytic expression
\begin{equation}
m_\text{crit} = \left( \sqrt{\frac{\pi}{2}} \,\frac{3 \hbar^2}{G \, a \, f(R/a) } \right)^{1/3}
\approx 5.153\times 10^{9}\,\mathrm{u}\,\left((a/\mu\mathrm{m}) \times f\left(\frac{R}{a}\right)\right)^{-1/3}
\end{equation}
for the critical mass. This expression is derived in \ref{app:comparison}.
The function $f$ is constantly 1 in case of the one-particle \SNE\ and shows an exponential
dependence on $R$ in case of the wide wave-function approximation. In case of the
Born-Oppenheimer approximation $f$ is a rather complicated function that can be found
in the appendix.

The resulting critical mass for a width of the centre-of-mass wave-function of
$0.5\,\mu\mathrm{m}$ is plotted as a function of the hollow-sphere radius in figure
\ref{fig:mcrit-hollow}.
The curve that the figure shows for the wide wave-function approximation coincides well
with the results we obtained in the purely numerical analysis in~\cite{Giulini.Grossardt:2013}.
For the Born-Oppenheimer-Type approximation the plot shows a radius dependence of the collapse mass
that is almost linear. This is in agreement with the result by Di\'osi~\cite{Diosi:1984} who estimates
the width of the ground state for a \emph{solid} sphere to be proportional to $(R/M)^{3/4}$.

\subsection{Narrow wave-functions in the Born-Oppenheimer scheme}
With the Born-Oppenheimer-Type approximation scheme just derived we now possess a tool
with which we can consider the opposite geometric 
situation than that in Case\,A, namely for widths of the 
centre-of-mass wave-function $\psi$ which are much smaller than 
the extensions (diameters of the support) of the matter 
distribution $\rob$, i.\,e., for well localised 
mass centres inside the bulk of matter.

Let us recall that in Newtonian gravitational physics 
the overall gravitational self-energy of a mass 
distribution $\rwf$ is given by 
\begin{equation}
\label{eq:GravSelfEnergy-1}
\Ug(\rwf):=
-\frac{G}{2}\int\,\D^3\vec x\int\,\D^3\vec x'\
\frac{\rwf(\vec x)\rwf(\vec x')}{\norm{\vec x-\vec x'}}\,.
\end{equation}
If $\rwf=\rho+\rho'$, we have by the simple quadratic 
dependence on $\rwf$ 
\begin{equation}
\label{eq:GravSelfEnergy-2}
\Ug(\rho+\rho'):=\Ug(\rho)+\Ug(\rho')+\Intg(\rho,\rho')\,,
\end{equation}
where 
\begin{equation}
\label{eq:GravSelfEnergy-3}
\Intg(\rho,\rho'):=
-G\int\,\D^3\vec x\int\,\D^3\vec x'\
\frac{\rho(\vec x)\rho'(\vec x')}{\norm{\vec x-\vec x'}}\,.
\end{equation}
represents the mutual gravitational \emph{interaction}
of the matter represented by $\rho$ with that represented 
by $\rho'$.  In the special case $\rho'=T_{\vec d}\rho$,
where $T_{\vec d}$ denotes the operation of translation by the
vector $\vec d$,
\begin{equation}
\label{eq:GravSelfEnergy-4}
\bigl(T_{\vec d}\rho\bigr)(\vec x):=\rho(\vec x-\vec d)\,,
\end{equation} 
we set
\begin{equation}
\label{eq:GravSelfEnergy-5}
I_\rho(\vec d):=\Intg(\rho,T_{\vec d}\rho)\,.
\end{equation} 

It is immediate from \eqref{eq:GravSelfEnergy-3} that 
$I_\rho:\reals^3\rightarrow\reals$ has a zero derivative 
at the origin $\vec 0\in\reals^3$, 
\begin{equation}
\label{eq:GravSelfEnergy-6}
I'_\rho(\vec 0)=0\,,
\end{equation}
and that it satisfies the following equivariance  
\begin{equation}
\label{eq:GravSelfEnergy-7}
I_\rho(R\vec d)=I_{\rho\circ R}(\vec d)
\end{equation} 
for any orthogonal $3\times 3$ matrix $R$. The latter 
implies the rather obvious result that the function 
$\vec d\mapsto I_\rho(\vec d)$ is rotationally invariant 
if $\rho$ is a rotationally invariant distribution, i.\,e., 
the interaction energy depends only on the modulus of the 
shift, not its direction. 

For example, given that $\rho$ is the matter density of a 
homogeneous sphere of radius $R$ and mass $M$,  
\begin{equation}
\label{eq:RhoHomSphere}
\rho(\vec x)=
\begin{cases}
\frac{3M}{4\pi R^3}&\text{for}\ \norm{\vec x}\leq R\\
0& \text{for}\ \norm{\vec x} > R\,,
\end{cases}
\end{equation}
the gravitational interaction energy is between two 
such identical distributions a distance $d:=\norm{\vec d}$
apart is 
\begin{equation}
I_\rho(d)=-\frac{GM^2}{R}\times
\begin{cases}
\frac{6}{5}-2\left(\frac{d}{2R}\right)^2+\frac{3}{2}\left(\frac{d}{2R}\right)^3-\frac{1}{5}\left(\frac{d}{2R}\right)^5
&\text{for}\ d\leq 2R\,,\\
\frac{R}{d}
&\text{for}\ d\geq 2R\,.
\end{cases}
\end{equation} 
The second line is obvious, whereas the first line follows, 
e.\,g., from specialising the more general formula (42) of 
\cite{Iwe:1982} to equal radii ($R_p=R_t$) and making the 
appropriate redefinitions in order to translate their 
electrostatic to our gravitational case. This formula 
also appears in~\cite{Colin.Durt.Willox:2014}.   
 
Using the definitions \eqref{eq:GravSelfEnergy-3} and \eqref{eq:GravSelfEnergy-5}, we can rewrite the right-hand 
side of \eqref{eq:born-oppenheimer} as convolution of 
$\abs{\psi}^2$ with $I_{\rob}$:
\begin{equation}
\label{eq:born-oppenheimer-rewirtten}
\Vg^{(B)}(t;\vec c)
=\int\,\D^3\vec c'\,
I_{\rob}(\vec c-\vec c')\,\abs{\psi(t;\vec c')}^2\,.
\end{equation}
Since in equation \eqref{eq:SNE} this potential is multiplied 
with $\psi(t,\vec c)$, we see that only those values of 
$I_{\rob}(\vec c-\vec c')$ will contribute where 
$\abs{\psi(t;\vec c')}^2\psi(t;\vec c)$ appreciably differs 
from zero. Hence if $\psi$ is concentrated in a region of 
diameter $D$ then we need to know $I_{\rob}(\vec c-\vec c')$
only for $\norm{\vec c-\vec c'}<D$. Assuming $D$ to be small 
we expand $I_{\rob}$ in a Taylor series. Because of 
\eqref{eq:GravSelfEnergy-6} there is no linear term, so 
that up to and including the quadratic terms we have (using 
that $\abs{\psi(t;\vec c')}^2$ is normalised with respect
to the measure $\D^3\vec c'$) 
\begin{equation}
\label{eq:born-oppenheimer-expanded-1}
\Vg^{(B)}(t;\vec c)
\approx I_{\rob}(\vec 0)+\tfrac{1}{2}I''_{\rob}(\vec 0)\cdot
\Bigl(
\vec c\otimes\vec c-
2\,\vec c\otimes \langle\vec c\rangle+
\langle\vec c\otimes\vec c\rangle
\Bigr)\,.
\end{equation}
Here $I''_{\rob}(\vec 0)$ denotes the second derivative of 
the function $I_{\rob}:\reals^3\rightarrow\reals$ at 
$\vec 0\in\reals$ (which is a symmetric bilinear form on 
$\reals^3$) and $\langle\,\cdot\,\rangle$
denotes the expectation value with respect to $\psi$. We stress 
that the non-linearity in $\psi$ is now entirely encoded into 
this state dependence of the expectation values which appear 
in the potential. If, for simplicity, we only consider 
centre-of-mass motions in one dimension, the latter being 
coordinatised by $c\in \reals$, then 
\eqref{eq:born-oppenheimer-expanded-1} simplifies to   
\begin{subequations}
\label{eq:born-oppenheimer-expanded-2}
\begin{alignat}{1}
\label{eq:born-oppenheimer-expanded-2a}
\Vg^{(B)}(t;c)
&\,\approx\, I_{\rob}(0)+\tfrac{1}{2}I''_{\rob}(0)
\Bigl(c^2-2c\,\langle c\rangle+
\langle c^2\rangle\Bigr)\\
\label{eq:born-oppenheimer-expanded-2b}
&\,=\,I_{\rob}(0)
+\tfrac{1}{2}I''_{\rob}(0)\bigl(c-\langle c\rangle\bigr)^2
+\tfrac{1}{2}I''_{\rob}(0)\bigl(\langle c^2\rangle-\langle c\rangle^2\bigr)\,.
\end{alignat}
\end{subequations}
The first term, $I_{\rob}(0)$, just adds a constant to the potential 
which can be absorbed by adding $-(\mathrm{i}/\hbar)I_{\rob}(0)t$ to the phase 
$\psi$. The second term is the crucial one and has been shown in 
\cite{Yang.etal:2013} to give rise to interesting and potentially 
observable for Gaussian states. 

More precisely, consider a one-dimensional non-linear 
Schr\"odinger evolution of the form \eqref{eq:SNE-Sys-a}
with $\Vg$ given by the second term in 
\eqref{eq:born-oppenheimer-expanded-2} and an additional 
external harmonic potential for the centre of mass, then 
we get the following non-linear Schr\"odinger-Newton 
equation for the centre-of-mass wave-function,
\begin{equation}
\label{eq:OneDimSNE-ExtPot}
\mathrm{i}\hbar\partial_t\psi(t;c)
=\left(-\frac{\hbar^2}{2M}\frac{\partial^2}{\partial c^2}
+\tfrac{1}{2}M\omega^2_cc^2
+\tfrac{1}{2}M\omega^2_{\rm SN}\bigl(c-\langle c\rangle\bigr)^2\right)\,\psi(t;c)\,,
\end{equation}
where $\omega_{\rm SN}:=\sqrt{I_{\rob}''(0)/M}$ is called the 
Schr\"odinger-Newton frequency. This equation has been 
considered in \cite{Yang.etal:2013}, where the last term 
on the right-hand side of \eqref{eq:born-oppenheimer-expanded-2b}
has been neglected for a priori no good reason. Note that 
$\langle c^2\rangle$ and $\langle c\rangle^2$ contain the 
wave function and hence are therefore not constant (in time). 
Now, in the context of \cite{Yang.etal:2013} 
the consequences of interest were the evolution equations 
for the first and second moments in the canonical phase-space 
variables, and it shows that for them only spatial derivatives
of the potential contribute. As a consequence, the term in question 
makes no difference. The relevant steps in the computation are 
displayed in \ref{app:moments}.

Based on the observation that equation \eqref{eq:OneDimSNE-ExtPot} 
evolves Gaussian states into Gaussian states, it has then been shown 
that the covariance ellipse of the Gaussian state rotates at frequency 
$\omega_q:=\sqrt{\omega_c^2+\omega^2_{\rm SN}}$ whereas the centre
of the ellipse orbits the origin in phase with frequency
$\omega_c$. This asynchrony results from a difference 
between first- and second-moment evolution and is a genuine 
effect of self gravity. It has been suggested that it may be 
observable via the output spectra of optomechanical
systems~\cite{Yang.etal:2013}.

\section{Conclusions and outlook}
Although the many-particle \SNE~\eqref{eq:n-particleSN} 
does not exactly separate into centre-of-mass and relative 
motion, we could show that for some well-motivated 
approximations such a separation is possible. As long as 
the extent of an object is negligible in comparison to 
the uncertainty in localisation of its centre of mass 
the one-particle equation \eqref{eq:SNE} is a good 
model in both approximation schemes considered.

In the opposite case of a well localised object, i.\,e. 
one that has a narrow wave-function compared to its 
extent, the gravitational potential takes the form
\eqref{eq:born-oppenheimer-expanded-2} which yields 
a closed system of equations for the first and second 
moments and therefore the effects described 
in~\cite{Yang.etal:2013}. The non-linear Schr\"odinger 
equation resulting from the potential 
\eqref{eq:born-oppenheimer-expanded-2} is also
considered in~\cite{Colin.Durt.Willox:2014}, where 
it is used for comparison of Schr\"odinger-Newton 
dynamics with models of quantum state reduction and 
decoherence.

The modification \eqref{eq:large-wf} provides a valid 
correction of the one-particle \SNE\ for objects of 
finite but small radii. This equation was considered 
in \cite{Jaeaeskelaeinen:2012} and studied numerically 
in \cite{Giulini.Grossardt:2013}. It remains unclear 
for which ratio of the object's extent to the width 
of the wave-function the Born-Oppenheimer-Type 
approximation~\eqref{eq:born-oppenheimer} starts to 
be superior to the wide wave-function approximation. 
It may even be the better approximation throughout 
the whole range of possible object sizes and 
wave-functions since a Born-Oppenheimer like
approximation is implicitly assumed also for the 
wide wave-function when the mass density is
taken to be that of a solid object.

In passing we make the final technical remark that the 
analysis of the critical mass for the hollow sphere shows 
that this mass increases linearly with the radius $R$ 
of the sphere. Given a fixed mass, this implies 
that the width of the stationary solution increases 
like $R^{3/4}$, a relation already found by 
Di\'osi~\cite{Diosi:1984}.

The interface between Quantum Mechanics and gravity theory 
remains one of the most interesting and profound challenges 
with hopefully revealing experimental consequences, which we 
are only beginning to explore. In this context one should also 
mention that non-linear one-particle Schr\"odinger equations 
are of course also considered for Einstein-Bose condensates, 
in which case inclusion of self gravity adds a Schr\"odinger-Newton 
term in addition to that non-linear term obtained from the 
effective potential within the Hartree-Fock approximation 
(Gross-Pitaevskii-Newton equation). Such equations are   
derivable for particle numbers $N\rightarrow\infty$ without 
further hypotheses and may open up the possibility to 
test self-gravity effects on large quantum systems. 
Recent experiments have demonstrated the high potential 
of atom interferometry on freely falling Einstein-Bose 
condensates~\cite{Muentinga.etal:2013} and it seems an 
interesting question whether this may be used to see 
self-gravity effects on such systems. 

\section*{Acknowledgements}
We gratefully acknowledge funding and support through the Center for
Quantum Engineering and Space-Time Research ({\small{QUEST}})
at the Leibniz University Hannover and the Center of Applied Space
Technology and Microgravity ({\small{ZARM}}) at the University of Bremen.
AG is supported by the John Templeton foundation (grant 39530).

\appendix

\section{Comparison of approximations for 
spherically symmetric mass distributions} \label{app:comparison}
For both the wide wave-function approximation \eqref{eq:large-wf} and the
Born-Oppenheimer-Type approximation \eqref{eq:born-oppenheimer}
one must solve integrals of the type
\begin{equation}
I(\vec a) = \int\D^3\vec r \; 
\frac{\rob(\vec r)}{\norm{\vec r - \vec a}} \,.
\end{equation}
In a spherically symmetric situation these take the form
\begin{align}
I(a) &= \int_0^\infty r^2 \, \D r \int_{-1}^1 \D \cos \theta \int_0^{2 \pi} \D \varphi \,
\frac{\rob(r)}{\sqrt{r^2 + a^2 - 2 r a \cos \theta}} \nonumber\\
&= \frac{4 \pi}{a} \int_0^a \D r \, r^2 \, \rob(r) + 4 \pi \int_a^\infty \D r \, r \, \rob(r) \,,
\end{align}
where we write $a$ for the absolute value $\abs{\vec a}$, etc.
If now we assume that $\rob$ is the mass density of a hollow 
sphere of radius $R$, i.\,e.
\begin{equation}
\rob(r) = \frac{M}{4 \pi \, r^2} \, \delta(r-R) \,,
\end{equation}
these integrals simplify to
\begin{equation}
I_R(a) = \begin{cases} \frac{M}{R} &\mbox{if } a < R \\
\frac{M}{a} & \mbox{if } a \geq R \end{cases} \,.
\end{equation}
With this the wide wave-function approximation \eqref{eq:large-wf} 
results in
\begin{equation}
\Vg^{(A)}(t;c;R) = -G M \int \D^3 \vec c' \, \abs{\psi(t;\vec c')}^2 \, I_R(\norm{\vec c - \vec c'})\,.
\end{equation}
On the other hand, the Born-Oppenheimer approximation \eqref{eq:born-oppenheimer} 
leads to 
\begin{equation}
\Vg^{(B)}(t;c;R) = -G \int \D^3 \vec c' \, \abs{\psi(t;\vec c')}^2 \int \D^3 \vec r' \,
\rob(\vec r') \, I_R(\norm{\vec c - \vec c' - \vec r'}) \,.
\end{equation}

In order to be able to obtain an analytical result we consider the initial Gaussian
wave packet
\begin{equation}
\label{eq:inital-gaussian}
\psi(t=0;\vec{c}) = (\pi a^2)^{-3/4} \, \exp\left(-\frac{c^2}{2a^2}\right) \,,
\end{equation}
for which these potentials take the form
\begin{align}
\label{eq:pot-initial-a}
V_A^0(c;R) &= -\frac{GM^2}{2} \Bigg\{ \frac{a}{\sqrt{\pi} \, c \, R}
\left[\exp\left(-\frac{(c+R)^2}{a^2}\right)-\exp\left(-\frac{(c-R)^2}{a^2}\right)\right]\nonumber\\
&\bleq +\frac{1}{c} \left[\erf\left(\frac{c+R}{a}\right)+\erf\left(\frac{c-R}{a}\right)\right]\nonumber\\
&\bleq +\frac{1}{R} \left[\erf\left(\frac{c+R}{a}\right)-\erf\left(\frac{c-R}{a}\right)\right]\Bigg\}\\
\label{eq:pot-initial-b}
V_B^0(c;R) &= -\frac{GM^2}{2} \Bigg\{ \frac{a}{\sqrt{\pi} \, c} \left(\frac{1}{R}+\frac{8 R}{3 a^2}\right)
\exp\left(-\frac{(c+2R)^2}{a^2}\right)\nonumber\\
&\bleq +\frac{a}{\sqrt{\pi} \, R} \left(\frac{1}{2c}-\frac{1}{4R}\right)
\left[\exp\left(-\frac{(c+2R)^2}{a^2}\right)-\exp\left(-\frac{(c-2R)^2}{a^2}\right)\right]\nonumber\\
&\bleq +\frac{1}{c} \left[\erf\left(\frac{c+2R}{a}\right)+\erf\left(\frac{c-2R}{a}\right)\right]\nonumber\\
&\bleq +\frac{1}{R} \left(1-\frac{c}{4R}-\frac{a^2}{8cR}\right)
\left[\erf\left(\frac{c+2R}{a}\right)-\erf\left(\frac{c-2R}{a}\right)\right]\Bigg\} \,.
\end{align}
Note that both potentials agree in the limits
\begin{equation}
\lim_{R\to0} V_{A,B}^0(c;R) = -\frac{GM^2}{c} \, \erf\left(\frac{c}{a}\right)
\quad \text{and} \quad \lim_{R\to\infty} V_{A,B}^0(c;R) = 0 \,.
\end{equation}

As a measure to compare these potentials with each other and the one-particle \SNE\
we use the second moment $Q(t) = \int \D^3 \vec{c} \, \abs{\vec c}^2 \abs{\psi(t;\vec{c})}^2$.
For a real wave packet its first order time derivative can be shown to vanish. Therefore the sign
of the second order time derivative $\ddot{Q}$ at $t=0$ determines if a wave packet initially
shrinks or increases in width. In general the second order time derivative is
\begin{equation}
\ddot{Q}(t) = \int \D^3 \vec{c} \left( \frac{2\hbar^2}{M^2} \abs{\vec{\nabla} \psi(t;\vec c)}^2
+ \frac{2}{M} \Vg(t;\vec c) \left( 3 \,\abs{\psi(t;\vec c)}^2
+ \vec c \cdot \vec \nabla \abs{\psi(t;\vec c)}^2 \right) \right)
\end{equation}
which for the spherically symmetric gaussian state \eqref{eq:inital-gaussian} takes the form
\begin{align}
\ddot{Q}(t=0) &= \frac{3 \hbar^2}{M^2 a^2} + \frac{8}{\sqrt{\pi} \,M\,a^5}
\int_0^\infty \D c \, \exp\left(-\frac{c^2}{a^2}\right) \, V^0(c) \, \left(3 a^2 c^2 - 2 c^4\right)\\
&= \frac{3 \hbar^2}{M^2 a^2} - \sqrt{\frac{2}{\pi}} \frac{G\,M}{a}\,f\left(\frac{R}{a}\right) \,.
\end{align}
The critical mass defined by $\ddot{Q}(t=0) = 0$ is then given by
\begin{equation}
m_\text{crit} = \left( \sqrt{\frac{\pi}{2}} \,\frac{3 \hbar^2}{G \, a \, f(R/a) } \right)^{1/3}
\approx 5.153\times 10^{9}\,\mathrm{u}\,\left((a/\mu\mathrm{m}) \times f\left(\frac{R}{a}\right)\right)^{-1/3} \,.
\end{equation}
The function $f$ is $f \equiv 1$ for the one-particle \SNE. For the hollow sphere
potential in the wide wave-function and Born-Oppenheimer-Type approximations,
\eqref{eq:pot-initial-a} and \eqref{eq:pot-initial-b}, respectively, this function can be calculated as
\begin{align}
f_A\left(\frac{R}{a}\right) &= \exp\left(-\frac{R^2}{2 a^2}\right) \\
f_B\left(\frac{R}{a}\right) &= \frac{2}{3} \sqrt{\frac{2}{\pi}} \exp\left(-\frac{4 R^2}{a^2}\right) \frac{R}{a}
\left(1-\left(\frac{2R}{a}\right)^2\right) \nonumber\\
&\bleq + \exp\left(-\frac{2 R^2}{a^2}\right) \left(1-\erf\left(\sqrt{2}\frac{R}{a}\right)\right)
\left(1+\frac{1}{3}\left(\frac{2R}{a}\right)^4\right) \nonumber\\
&\bleq +\frac{a^2}{2 R^2} \left(\frac{1}{\sqrt{2}}\erf\left(2\frac{R}{a}\right)
- \exp\left(-\frac{2 R^2}{a^2}\right)\erf\left(\sqrt{2}\frac{R}{a}\right)\right) \,.
\end{align}

\section{Divergence of the solid-sphere potential in the wide wave-function approximation}
\label{app:solid-sphere-divergence}
Given a spherically symmetric situation the wide wave-function approximation \eqref{eq:large-wf}
takes the form
\begin{equation}\label{eqn:app1}
V(c) = (\abs{\psi}^2 * \Phi)(c)
= 4 \pi\; \int_0^\infty \D c' \; c'^2 \, \abs{\psi(c')}^2 \, \Phi(\abs{c-c'}) \,,
\end{equation}
where for the potential $\Phi$ we want to consider the following three cases:
\begin{itemize}
\item Coulomb potential (i.\,e. the case of the \SNE \eqref{eq:SNE}):
\begin{equation}
\Phi_c(c) = -\frac{1}{c} \,,
\end{equation}
\item hollow sphere of radius $R$:
\begin{equation}
\Phi_h(c) = \begin{cases} -\frac{1}{R} &\mbox{if } c < R \\
\Phi_c(c) & \mbox{if } c \geq R \end{cases} \,,
\end{equation}
\item solid sphere of radius $R$:
\begin{equation}
\Phi_s(c) = \begin{cases} -\frac{3}{2R} + \frac{c^2}{2R^3} &\mbox{if } c < R \\
\Phi_c(c) & \mbox{if } c \geq R \end{cases} \,.
\end{equation}
\end{itemize}

First we want to study the behaviour of $V_0 = V(c=0)$ for a Gaussian wave packet
$\abs{\psi(c)}^2 = \frac{1}{4 \pi} \exp(-c^2)$. For convenience we omit all pre-factors.
Equation \eqref{eqn:app1} then reads:
\begin{align}
V_0 &= \int_0^\infty \D c \; c^2\,\exp(-c^2) \, \Phi(c) \\
&= \int_0^R \D c \; c^2\,\exp(-c^2) \, \Phi(c) + \int_R^\infty \D c \; c^2\,\exp(-c^2) \, \Phi_c(c) \\
&= \int_0^R \D c \; c^2\,\exp(-c^2) \, \Phi(c) -\frac{\exp(-R^2)}{2} \,.
\end{align}
For the three different potentials one obtains
{\allowdisplaybreaks%
\begin{align}
V_{0,c} &= \int_0^R \D c \; c^2\,\exp(-c^2) \, \Phi_c(c) -\frac{\exp(-R^2)}{2} \nonumber\\
&= -\frac{1}{2} +\frac{\exp(-R^2)}{2} -\frac{\exp(-R^2)}{2} \nonumber\\
&= -\frac{1}{2} \\
V_{0,h} &= \int_0^R \D c \; c^2\,\exp(-c^2) \, \Phi_h(c) -\frac{\exp(-R^2)}{2} \nonumber\\
&= \frac{\exp(-R^2)}{2} -\frac{\sqrt{\pi}}{4R} \erf(R) -\frac{\exp(-R^2)}{2} \nonumber\\
&=  -\frac{\sqrt{\pi}}{4R} \erf(R) \label{eqn:app-A9}\\
V_{0,s} &= \frac{3}{2} \int_0^R \D c \; c^2\,\exp(-c^2) \, \Phi_h(c)\nonumber\\
&\bleq + \frac{1}{2R^3} \int_0^R \D c \; c^4 \exp(-c^2) -\frac{\exp(-R^2)}{2} \nonumber\\
&= -\frac{3\sqrt{\pi}}{8R} \erf(R) -\frac{3}{8R^2} \exp(-R^2) -\frac{3\sqrt{\pi}}{16R^2} \erf(R)
\label{eqn:app-A10}
\end{align}}%
In the limit $R \to 0$ the function $\erf(R)/R$ converges to $2/\sqrt{\pi}$. Thus, \eqref{eqn:app-A9} converges
to $-1/2$ and yields the same value as one gets for $\Phi_c$. For \eqref{eqn:app-A10}
both the second and third term diverge but the sum of both terms converges and altogether
$V_{0,s}$ also converges to the value of $-1/2$. So everything seems fine.

But now consider the behaviour of $V(c)$ in a small neighbourhood of $c=0$, i.\,e.
$V_\eps = V(c = \eps)$. For the hollow sphere this changes nothing of course, since the
potential is constant within radius $R$. The potentials
$\Phi_c$ and $\Phi_s$ can be expanded around $\eps = 0$ and yield
\begin{align}
\Phi_c(c+\eps) &= \Phi_c(c) + \eps\,\frac{1}{c^2} + \mathcal{O}(\eps^2) \\
\Phi_s(c+\eps) &= \Phi_s(c) + \eps\,\frac{c}{R^3} + \mathcal{O}(\eps^2) \,.
\end{align}
This gives the additional contributions
\begin{align}
V_{\eps,c} &= V_{0,c} + \eps \,\int_0^R \D c \; \exp(-c^2) \\
&= V_{0,c} + \eps \, \frac{\sqrt{\pi}}{2} \, \erf(R) \\
V_{\eps,s} &= V_{0,s} + \frac{\eps}{R^3} \,\int_0^R \D c \; c^3 \,\exp(-c^2) \\
&= V_{0,s} - \eps \, \frac{1+R^2}{2} \, \exp(-R^2) + \frac{\eps}{2R^3}
\end{align}
to the potentials. For the Coulomb potential everything is fine since
$\erf(R) \to 0$ for $R \to 0$. Hence, both the Coulomb and the hollow sphere potential
obtain no further contributions at this order and it can be easily checked that
this also holds for all higher orders in $\eps$.

For the solid sphere potential, however, things are not fine at all. Not only does
the term proportional to $\exp(-R^2)$ in the limit $R \to 0$ yield a contribution
$-\eps/2$ which already makes it differ from the Coulomb potential. The last term
is even worse because it diverges in this limit. Therefore, we cannot take this model
seriously for small radii of the solid sphere and we are better off taking the hollow sphere potential
as a toy model for the density of a molecule.

\section{Evolution equations for first and second moments in the narrow wave-function limit}
\label{app:moments}
\newcommand{\erw}[1]{\langle #1 \rangle}
\newcommand{\x}{\vec{x}}
\newcommand{\xx}{\vec{x^2}}
\newcommand{\erwx}{\erw{\vec{x}}}
\newcommand{\erwxx}{\erw{\vec{x}^2}}
\newcommand{\p}{\vec{p}}
\newcommand{\pp}{\vec{p^2}}
\newcommand{\erwp}{\erw{\vec{p}}}
\newcommand{\erwpp}{\erw{\vec{p}^2}}
\newcommand{\erwxp}{\erw{\vec{x} \cdot \vec{p}}}
\newcommand{\erwpx}{\erw{\vec{p} \cdot \vec{x}}}
\newcommand{\ksn}{k_\text{SN}}
\newcommand{\kcm}{k_\text{CM}}
\newcommand{\iik}[1]{\int \D^3 x \left( #1 \right)}
\newcommand{\ixi}[1]{\int \D^3 x \, x_i \left( #1 \right)}
\newcommand{\ixj}[1]{\int \D^3 x \, x_j \left( #1 \right)}
\newcommand{\ii}{\int \D^3 x \,}
\newcommand{\pd}{\dot{\psi}}
\newcommand{\psq}{\abs{\psi}^2}
Here we will explicitly derive the self-contained system of evolution equations for
the first and second moments given in \cite{Yang.etal:2013}. It has been noted there that since this
system is closed, Gaussian states will remain Gaussian under evolution. We will show that the
difference of our equation \eqref{eq:born-oppenheimer-expanded-2} to equation \eqref{eq:OneDimSNE-ExtPot}
given in \cite{Yang.etal:2013} has no influence on this set of equations.

For this we consider the Schr\"odinger equation
\begin{equation}
 \mathrm{i} \dot{\psi} = \frac{p^2}{2 M} \psi + H_1 \psi \,,
\end{equation}
where
\begin{align*}
H_1 &= \frac{k}{2} \xx - \ksn \x \cdot \erwx + \alpha \erwx^2 + \beta \erwxx \\
p_i &= -\mathrm{i} \partial_i \\
k &= \kcm + \ksn = M \omega_\text{CM}^2 + M \omega_\text{SN}^2 \,.
\end{align*}
In principle, in the case of equation \eqref{eq:born-oppenheimer-expanded-2} we have $\alpha = 0$,
$\beta = \ksn / 2$, while in the case of equation \eqref{eq:OneDimSNE-ExtPot} $\alpha = \ksn / 2$,
$\beta = 0$. But note that
\begin{subequations}\begin{align}
\partial_i \erwx_j &= 0 \\
\partial_i \erwxx &= 0 \,.
\end{align}\end{subequations}
Therefore, independent of the choice of $\alpha$ and $\beta$ the derivatives of $H_1$ are
\begin{subequations}\begin{align}
\partial_i H_1 &= k x_i - \ksn \erwx_i \\
\Delta H_1 &= 3 k \,.
\end{align}\end{subequations}
We will see that $H_1$ will enter into the evolution equations for the first and second moments only
through these derivatives. Thus, for the different equations
\eqref{eq:born-oppenheimer-expanded-2} and \eqref{eq:OneDimSNE-ExtPot} we obtain the same
evolution equations for the first and second moments, which are:
\begin{subequations}\begin{align}
\partial_t \erwx_i
&= \ixi{ \psi^* \pd + \psi \pd^* } = \frac{\mathrm{i}}{2M} \ixi{ \psi^* \Delta \psi - \psi \Delta \psi^* } \\
&= \frac{\mathrm{i}}{2M} \iik{ -(\partial_j \psi) \partial_j(x_i \psi^*) + (\partial_j \psi^*) \partial_j(x_i \psi) } \\
&= \frac{\mathrm{i}}{2M} \iik{ -\psi^* \partial_i \psi + \psi \partial_i \psi^* } = \frac{1}{M} \ii{ \psi^* (-\mathrm{i}\partial_i) \psi } \\
&= \frac{\erwp_i}{M}
\end{align}\end{subequations}
\begin{subequations}\begin{align}
\partial_t \erwp_i
&= -\mathrm{i} \iik{ \pd^* \partial_i \psi + \psi^* \partial_i \pd } \\
&= \frac{1}{2M} \iik{-(\Delta \psi^*) \partial_i \psi + \psi^* \partial_i \Delta \psi} - \erw{\partial_i H_1} \\
&= \frac{1}{2M} \iik{(\partial_j \psi^*) \partial_i \partial_j \psi + \psi^* \partial_i \Delta \psi} - \erw{\partial_i H_1} \\
&= \frac{1}{2M} \iik{-\psi^* \partial_i \Delta \psi + \psi^* \partial_i \Delta \psi} - \erw{\partial_i H_1} \\
&= -k \erwx_i + \ksn \erwx_i \\
&= -\kcm \erwx_i
\end{align}\end{subequations}
\begin{subequations}\begin{align}
\partial_t \erwxx
&= \ii \vec{x}^2 \left( \psi^* \pd + \psi \pd^* \right)
= \frac{\mathrm{i}}{2M} \ii \vec{x}^2 \left( \psi^* \Delta \psi - \psi \Delta \psi^* \right) \\
&= \frac{\mathrm{i}}{2M} \iik{-(\partial_j \psi) \partial_j(x_i x_i \psi^*) + (\partial_j \psi^*) \partial_j (x_i x_i \psi)} \\
&= \frac{\mathrm{i}}{M} \iik{ -x_i \psi^* \partial_i \psi + x_i \psi \partial_i \psi^* } \\
&= \frac{1}{M} \iik{ \psi^* x_i (-\mathrm{i} \partial_i) \psi + \psi^* (-\mathrm{i} \partial_i) (x_i \psi) } \\
&= \frac{1}{M} \left( \erwxp + \erwpx \right)
\end{align}\end{subequations}
\begin{subequations}\begin{align}
\partial_t \erwpp
&= -\iik{\pd^* \Delta \psi + \psi^* \Delta \pd} 
= \frac{\mathrm{i}}{2M} \iik{(\Delta \psi^*) \Delta \psi - \psi^* \Delta \Delta \psi} \nonumber\\
&\hspace{14em} + 2 \mathrm{i} \ii \psi^* (\partial_i H_1) \partial_i \psi + \mathrm{i} \erw{\Delta H_1} \\
&= 2 \mathrm{i} \ii \psi^* \left( k x_i - \ksn \erwx_i \right) \partial_i \psi + 3 \mathrm{i} k \\
&= -2k \ii \psi^* x_i (-\mathrm{i} \partial_i) \psi + 2 \ksn \erwx_i \ii \psi^* (-\mathrm{i} \partial_i) \psi + 3 \mathrm{i} k \\
&= -2k \erwxp + 2 \ksn \erwx \cdot \erwp + k \erwxp - k \erwpx \\
&= - k \left( \erwxp + \erwpx \right) + 2 \ksn \erwx \cdot \erwp
\end{align}\end{subequations}
\begin{subequations}\begin{align}
\partial_t \erwxp
&= \partial_t \erwpx = \iik{\pd^* x_i (-\mathrm{i} \partial_i) \psi + \psi^* x_i (-\mathrm{i} \partial_i) \pd} \\
&= -\frac{1}{2M} \ixi{(\Delta \psi^*) \partial_i \psi - \psi^* \partial_i \Delta \psi} - \erw{x_i \partial_i H_1} \\
&= \frac{1}{2M} \ii (\partial_j \psi^*) \partial_j(x_i \partial_i \psi)  + \frac{1}{2M} \ii x_i \psi^* \partial_i \Delta \psi - \erw{x_i \partial_i H_1} \\
&= \frac{1}{2M} \iik{ (\partial_i \psi^*) \partial_i \psi + x_i (\partial_j \psi^*) \partial_i \partial_j \psi
 + x_i \psi^* \partial_i \Delta \psi} - \erw{x_i \partial_i H_1} \\
&= \frac{1}{2M} \iik{-\psi^* \Delta \psi - \psi^* \partial_j(x_i \partial_i \partial_j \psi) + x_i \psi^* \partial_i \Delta \psi} - \erw{x_i \partial_i H_1} \\
&= -\frac{1}{M} \ii \psi^* \Delta \psi - \erw{x_i \partial_i H_1} \\
&= \frac{\erwpp}{M} - k \erwxx + \ksn \erwx^2
\end{align}\end{subequations}

The same evolution equations are obtained by operators
$\vec x$ and $\vec p$ that in the Heisenberg picture fulfil
\begin{subequations}\begin{align}
\partial_t \vec x &= \frac{\vec p}{M} \\
\partial_t \vec p &= -\kcm \vec x - \ksn (\vec x - \erw{\vec x}) \,.
\end{align}\end{subequations}
This was used in \cite{Yang.etal:2013} to describe the effect of the \SNE\ on Gaussian states.


\section*{References}

\bibliographystyle{iopart-num}
\bibliography{giulini_grossardt_sn_center_of_mass}

\end{document}